# Shock machine for the mechanical behaviour of hip prostheses: a description of performance capabilities.

Juliana Uribe[1], Jean Geringer[1], Bernard Forest[1]
[1]Center for Health Engineering, Biomechanics and Biomaterials Department
UMR CNRS 5146, SFR Ifresis
Ecole Nationale Supérieure des Mines de Saint-Etienne,
158 cours Fauriel, 42023 Saint-Etienne, France
Corresponding author: *uribe@emse.fr*

**ABSTRACT**

The aim of this study is to describe the behaviour of a shock machine designed for testing hip prostheses. A micro-separation between head and cup occurs inducing a shock of several times the body weight, leading to fracture of ceramic femoral components. Femoral heads and cups of diameter 32mm manufactured from alumina were tested in dry and wet conditions. Implants were subjected to shocks with a load-profile of 9kN load at 2Hz and various microseparations. Position is monitored and force is measured with two acquisition systems. The working range and the device capabilities are investigated. Only cups tested in dry conditions failed. Observations by Scanning Electron Microscopy revealed intergranular and transgranular fractures. Two wear stripes are observed on the heads. Three-dimensional roughness of wear stripes is measured. Since experimental results are in good agreement with retrieved femoral heads, the shocks machine reproduces the *in vivo* degradations.

## INTRODUCTION

Total hip arthroplasty consists of replacing hip joint with an artificial femoral head, cup and acetabular shell, Figure 1a. Because of the increasing life expectancy, there is a need of prostheses with a longer lifetime. Ceramics are submitted to lower friction and more reduced wear than the ones of metal or polymer (Ultra High Molecular Weight PolyEthylene) bearing surfaces. [1,2] Aseptic loosening with accompanying osteolysis is less frequent for ceramics cups compared to polymer ones, therefore better long term results are obtained. However, ceramics could be worn and, in some cases fractured, when subjected to shocks.

To reproduce *in vitro* the *in vivo* wear, a description of the hip joint biomechanics is needed. Normal gait cycle has two force peaks. After hip replacement, a microseparation between cup and head could occur. Thus, when heel hits ground, a shock is produced with a vertical force[3,4], Figures 1b and 2, up to 600% body weight [5], which means approximately a force of 9 kN for a patient who weights 150 kg.

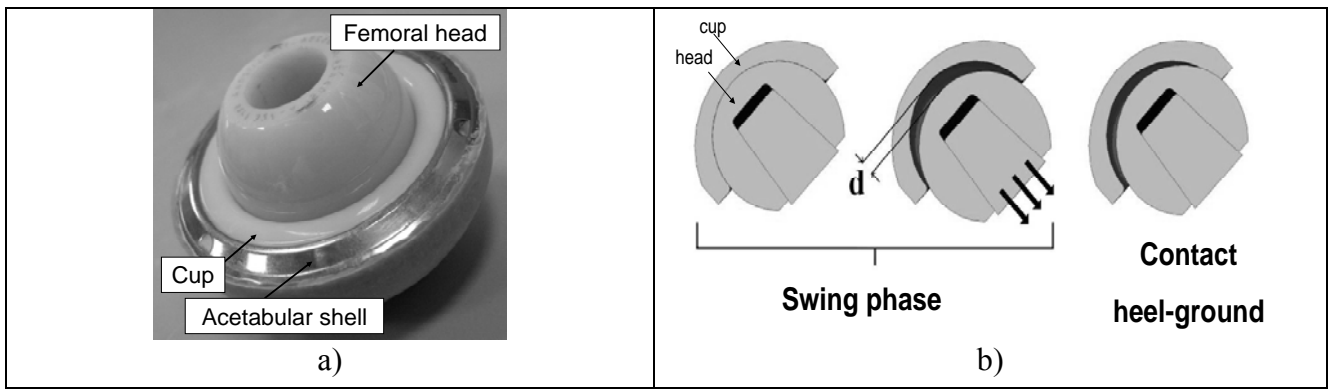

Figure 1. a) Hip prosthesis and b) Microseparation phenomena, d represents the distance of microseparation[3,4].

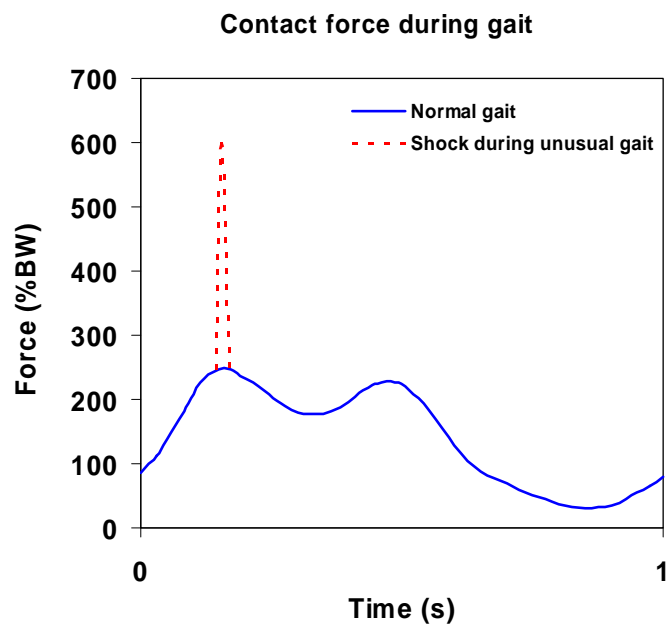

Figure 2. Hip contact force in %BW. Gait cycle showing the maximum of force when the heel hits the ground, BW: Body Weight[5].

According to the ISO 14242-1, hip implants should be tested before implantation with a hip simulator. This device is used in order to measure the wear rate of femoral components with the aim of establishing the limits of performance for the assembly cup-head. The cup and the head are in contact, so no microseparation is possible and the shock influence is not studied. In the last decade some investigations were in progress to study shocks in hip simulators[6-9]. This point is the most relevant for ceramic hip implants because it induces degradations on the contrary to the usual hip walking simulator.

Thus, the shock machine, considered in this work, is a device studying the influence of imposed microseparation in the degradation of artificial hip joint, head/cup assembly. The microseparation is monitored precisely with this device.

## MATERIALS AND METHODS

The shock machine (QUIRI Hydromecanique) is shown in Figure 3. This device was designed specially for testing hip prostheses, head and cup assembly, subjected to shocks.

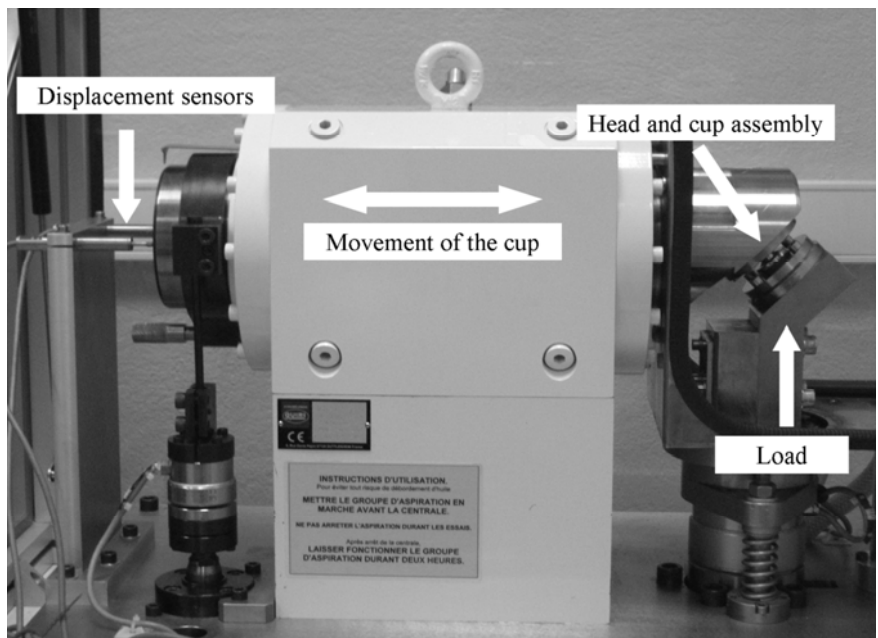

Figure 3. Shock machine

Force varies between from 0 to 10kN with an adjustable duration from 20 to 100ms and a frequency between 0.1 and 5Hz. Tests in both dry and wet conditions are possible. The main feature that makes this machine different from usual hip simulators is the opportunity to impose a microseparation. A control by displacement is possible. Microseparation between cup and head was controlled.

Cups and heads were manufactured from alumina of diameter 32mm. The radial clearance between the head and the cup is ranging between 50 and 100µm. The actual value of radial clearance will not be exposed due to confidential agreement with the manufacturer (SERF company). Grain size, density and Young's modulus were in agreement with the ISO 6474-1 standard requirements, Table I. The ceramic cups were sheathed in an acetabular shell. In order to accelerate degradations, the tests were carried out in severe conditions; no lubrication and a 30ms load of 9kN at 2Hz[4]. Assembly was inclined at 45° to respect standard anatomic position, Figure 4. The cup was fixed and the head was maintained into a cone.

Table I. Properties of tested materials; the tests were carried out according to ISO 6474-1

|  | Grain size (μm) | Density (g.cm$^{-3}$) | Young's modulus (GPa) |
|---|---|---|---|
| Head | 1.49 | 3.97 | 402.0 |
| Cup | 1.50 | 3.97 | 404.5 |

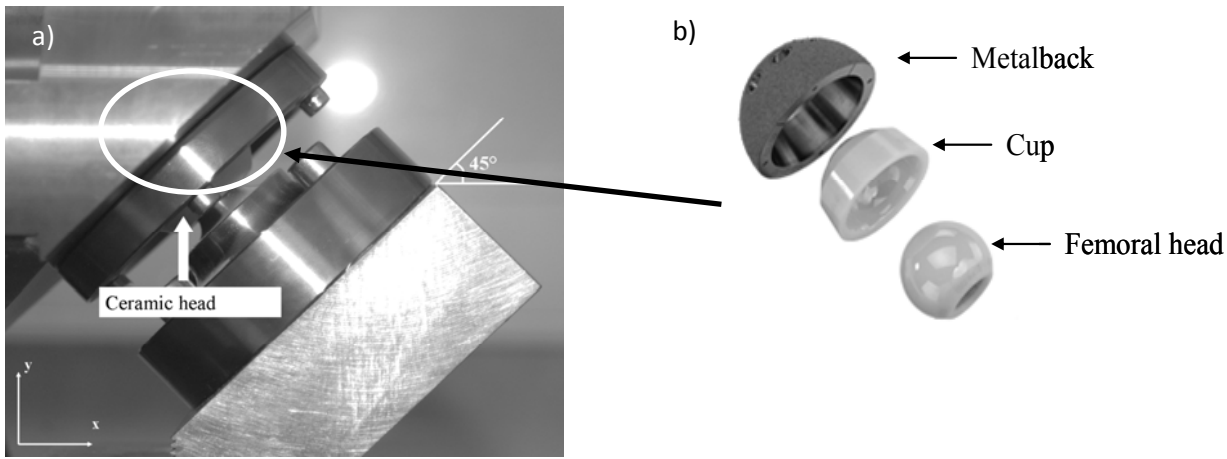

Figure 4. a) Head and cup assembly inclined at 45°; b) head/cup/metal back assembly

*Determining the behaviour of the cup after impact*

As mentioned above, the cup is fixed to a support made from 316L Stainless Steel, with only one degree of freedom: horizontal x-axis. In order to determine the x-displacement of the cup after shocking, two inductive displacement transducers, LVDT (HBM W5TK), were set at the left side, at the end of the support as shown in Figure 3. The displacement signals of the cup were then recorded during 20s for a load of 9kN at 2Hz. Three types of loading signal were studied: square, sinus and walking. Tension signals were recorded using LabView® 8.5 software with a data acquisition system (Ni-DAQ). Sampling frequency was set to 1 kHz. A code was developed in Matlab® for postprocessing.

*Machine validation protocol*

The machine software allows applying 3 displacement signals: sine, triangle and user's defined signal. Sine wave is the most easy to apply because its form without peaks. Since triangle is a signal with two peaks and an abrupt change in slope, the control is less performing. User's defined signal is a signal with a peak which reproduces the shock during gait cycle.

Several values of microseparations (d), Figure 5, were tested: 0, 0.7, 1, 1.3, 1.6 and 1.9mm. They are related to the *in vivo* behaviour. To each microseparation corresponded a vertical distance, (dv). For each microseparation, a displacement of X was imposed to the head. Since the distance between the head and the cup was dv, a force is applied when X>dv and increased with X. The distance g = X-dv was defined as "relative distance", Figure 5.

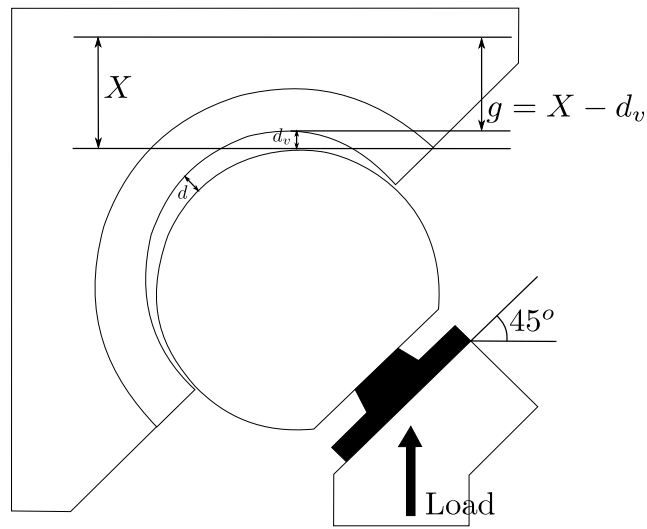

Figure 5. Virtual distance g=X-dv. d: microseparation, dv: vertical distance to obtain the value d. X is the displacement imposed to the head

For each microseparation, d, the relative distance, g, was increased by 0.5mm. All test conditions are shown in Table II. In order to compare raw data to oscilloscope signals provided by the electronic elements of the device, both force signals were recorded simultaneously during 20s with an acquisition system (LabView® 8.5). The average force was also calculated.

Table II. Parameters for validating the shock machine. Filled dot, ●, represents sinus and triangle signals, and empty dot, '○', represents the signal describing the shock during gait cycle.

| g(mm)<br>d(mm) | 0.5 | 1.0 | 1.5 | 2.0 | 2.5 | 3.0 | 3.5 | 4.0 | 4.5 | **5.0** |
|---|---|---|---|---|---|---|---|---|---|---|
| 0 | ● ○ | ● ○ | | | | | | | | |
| 0.75 | ● ○ | ● ○ | ○ | ○ | ○ | ○ | | | | |
| 1 | ● ○ | ● ○ | ● ○ | ○ | ○ | ○ | ○ | | | |
| 1.3 | ● ○ | ● ○ | ● ○ | ○ | ○ | ○ | ○ | ○ | ○ | |
| **1.6** | ● ○ | ● ○ | ● ○ | ○ | ○ | ○ | ○ | ○ | ○ | ○ |

*Method for modelling the behaviour of the cup*

Displacements signals from LVDT, see Figure 3, were recorded and analysed in order to determine the mechanical behaviour of the cup after the impact. The cup response was described as a damped spring-mass system. The cup oscillates during 0.2s which is about seven times the duration of the peak of force. Positive sign means the cup moves in the positive x direction.

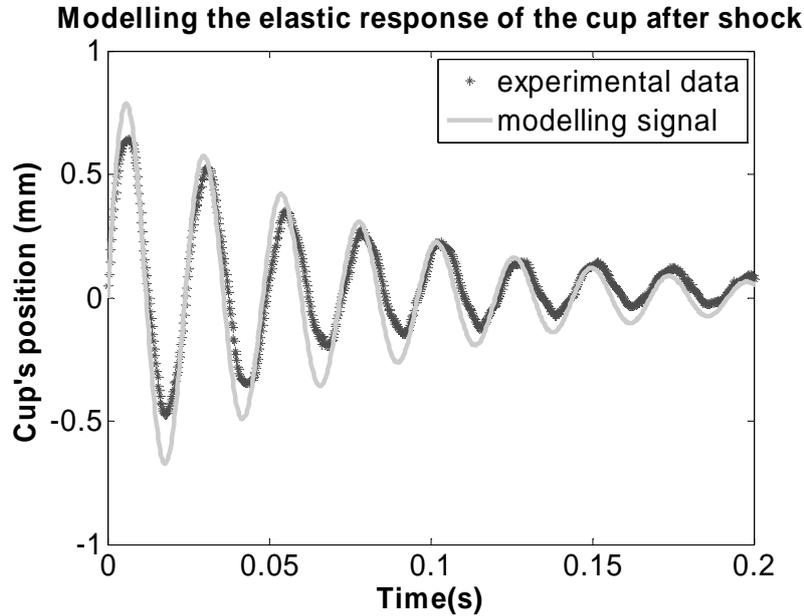

1 espace tim e ?

Figure 6. Results for a square signal, with a microseparation of 1.3mm and a force of 9kN, black dotted line: the elastic response of the cup after the shock, gray continued line: the modelling signal.

The process of data acquisition, processing and modelling is summarized in Figure 7. Spring constant, *k*, and damping coefficient, *c*, were determined by fitting using Matlab®.

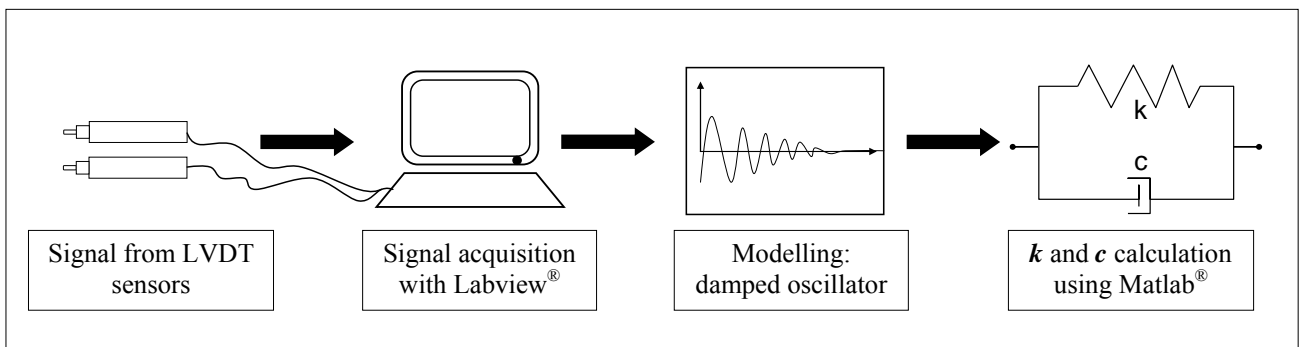

Figure 7. Diagram of data acquisition and postprocessing

The damped harmonic oscillator equation, describing the movement of the cup after the shock is supposed equal to:

$$\frac{d^2x}{dt^2} + 2\zeta\omega_0\frac{dx}{dt} + \omega_0^2 x = 0 \qquad (1)$$

where x is the displacement of the cup in the axis perpendicular to the applied load, $\zeta$ is a constant called the damping ratio and $\omega_0$ is the undamped angular frequency. For a mass, m, the elastic constant *k* and the damping coefficient *c* are given by:

$$k = \omega_0^2 * m \tag{2}$$

and

$$c = 2 * \zeta * \omega_d * m \tag{3}$$

where $\omega_d$ is the damped natural frequency. The solution to this underdamped harmonic oscillator is an exponentially decaying sinusoid:

$$f(t) = A * \sin(\omega_d * t) * e^{-\zeta \omega_0 t} \quad \text{where} \quad \omega_d = \omega_0 \sqrt{1-\zeta^2} \tag{4}$$

and A is the amplitude.

Since the response of the cup was similar regardless the signal studied, we decided calculating the parameters *k* and *c* using only one type of signal. As an example, the parameters were calculated using Matlab® for a square signal, with a load of 9kN and a microsepration of 1.3mm.

$$A = (0.65 \pm 0.02) mm \tag{5}$$
$$\zeta = 0.050 \pm 0.001 \tag{6}$$
$$\omega_d = (262.00 \pm 0.01) rad \cdot s^{-1} \tag{7}$$

The mass of the cup support was m=45kg. The parameters *k* and *c* were then calculated as:

$$k = 3.1 x 10^6 \, N \cdot m^{-1} \tag{8}$$
$$c = 1.1 x 10^3 \, Kg \cdot s^{-1} \tag{9}$$

As can be seen in Figure 6, the underdamped oscillator model fits experimental data well. The *k* and *c* values will be adopted in a finite element modelling, further investigations not mentioned in this work, for reproducing the movement of the cup after a shock and studying stresses distribution on the assembly cup-head.
Finally, the shocks device could be described in a more simplified manner as shown in Figure 8.

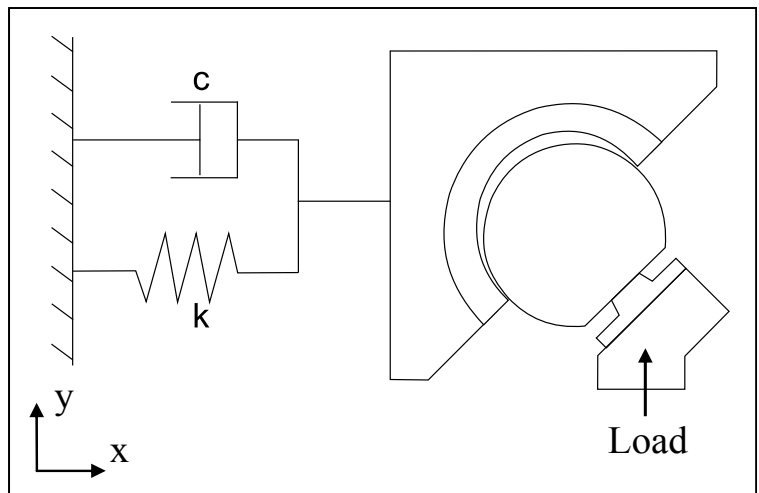

Figure 8. Scheme of shocks machine. The movements of the cup and the device are represented by a damped spring system.

*Relation between Force and relative distance, g*

The validation tests were made to determine the range of operating conditions to reach a force of 9kN, and 3 signals were studied: sinus, triangle and walking. The signals from Ni-DAQ$^{TM}$ and machine software, Quiri$^{TM}$ did not show significant difference for sinus and triangle. On the other hand, for walking signal, recorded signals were considerably different for forces under 2kN. Above 2kN, signals are similar and may be modelled by a linear curve which slope is calculated for each microseparation value (Figure 9 and Table III).

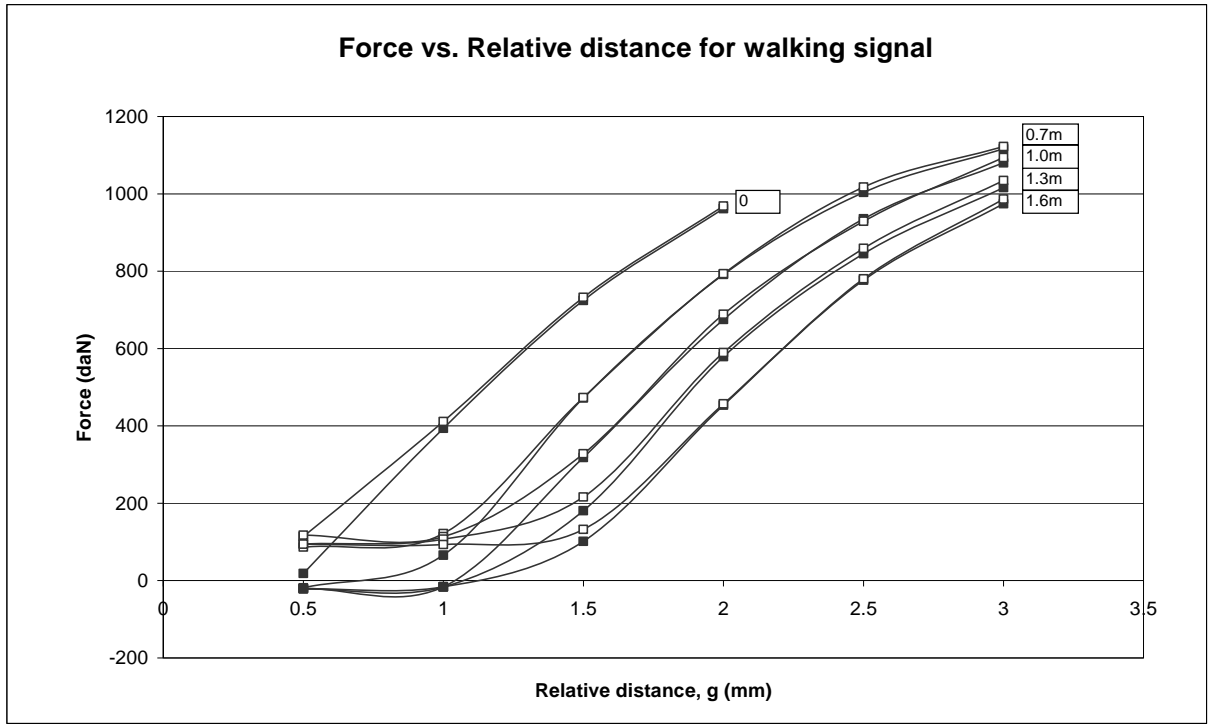

Figure 9. Force on the assembly cup-head as a function of relative distance of the head, for different values of microseparation; □: Ni-DAQ, ■: machine software

*Problème avec les unites, mm*
*J'enlèverais les lignes sur le graphe*
*J'effacerais le -200*

Table III. Slope of linear sections for tested microseparations

| Microseparation (mm) | Slope of the linear section (daN/mm) | | $\frac{Ni - DAQ - Machine}{Ni - DAQ} x100\%$ |
|---|---|---|---|
| | Machine | Ni-DAQ | |
| 0.0 | 632 | 576 | 9.6 |
| 0.7 | 429 | 434 | 1.2 |
| 1.0 | 509 | 507 | 0.3 |
| 1.3 | 554 | 544 | 1.8 |
| 1.6 | 588 | 577 | 1.9 |

Tests had duration of 20s and were stopped when force was greater than 11kN to avoid fracture of ceramic components and to be consistent with device capabilities. The results in Figure 9 show that a relative distance of 1.8mm has to be imposed to the head for reaching 9kN, without microseparation. Hence, from Figure 9, an equation was established for microseparation of 1.3mm, the typical microseparation for *in vivo* tests:

$$F = (549.6 \pm 6.8) * g - (571.3 \pm 29.0) \qquad (10)$$

where F is the desired force in daN and g is the relative distance to be imposed in mm.

From Figure 8 and Equation 10, a relative distance of 2.7mm is needed for reaching a force of 9kN. Following investigations for standard testing will be considered with microseparation [3] of 1.3mm and g=2.7mm. Finally the shock device is completely characterized for applying the closest conditions to the actual movement of the head-cup assembly.

*Tests procedure*

Materials for test were six heads and cups made from alumina, diameter of 32mm. Only one couple was investigated for the device calibration. After testing all microseparations, the standard value d=1.3mm and required virtual distance g, were taken for an extended test. The assembly was inclined at 45°. Loading conditions were defined taking into account the study of gait cycle which shows that a shock could occur with a peak of force up to six times the body weight[5]. It means that for a patient with an artificial joint who weighs 150kg, a force of 9kN is applied to the assembly cup-head. The frequency was set to 2Hz in order to accelerate the test.
In physiological normal conditions, hip joint is lubricated by the synovial fluid, which is made of hyaluronic acid, lubricin, proteinases, collagenases and white blood cells[10]. Lubrication may vary from patient to patient depending on the condition of the surrounding tissues. We study two extreme cases: dry and wet conditions. As recommended by ISO 14242-1 about the environmental conditions for wet test, newborn calf serum (17.5 g.L$^{-1}$ and 1g.L$^{-1}$ sodium azide) was used as lubricant. It contains a mixture of albumins, Igs, growth factors, minerals, vitamins and some complex lipids. Calf serum was changed weekly for avoiding proteins degradations by oxidation that could lead to different lubrication conditions. Temperature remained stable at 36°C ± 2. Tests were carried out during 800,000 cycles when no fracture was observed.

Heads and cups were observed and photos were taken every 50,000 cycles in order to determine the degree of damage and measurements of the width of the wear stripe were taken.

Roughness of femoral head tested in dry conditions was measured by AFM (Digital Instruments, multimode AFM, SiN probe, stiffness of 0.12N/m, Software 6.13). A particular attention was paid on the manufacturing process of the AFM samples from the ceramic head. Since the maximum size of samples is 1mm width and 0.5mm height, all heads were carefully sawed, without embrittlement of worn zones. The analyzed area by AFM is 40x40 µm$^2$. Five AFM measurements were carried out in both unworn and worn surfaces. A first order flatten filter was applied after acquisition in order to remove noise and tilt. The polynomial $z = a + bx$ is subtracted from each scan line, removing offset (a) and slope (b).

## RESULTS

*Investigating the lifetime of head-cup assembly*

Six heads and cups manufactured from alumina, diameter 32mm, were tested: three in dry conditions and three using a calf serum as lubricant. The load was 9kN with an initial microseparation of 1.3mm. From Figure 9, a virtual distance of g=2.7mm was imposed. After characterizing the machine, this part aims at analyzing two typical wear zones on the head.

*Cup*

Tests and cycles at fracture are summarized in Table IV. The three cups from dry tests fractured at the lower rim of the cup, Figure 10. Cup fracture occurred at 254,000 ± 43,000 cycles. Fractured bearing surfaces observed by Scanning Electron Microscopy, SEM, revealed intergranular and transgranular fractures,Figure 11. The size of debris from fractured cups was less than 1mm to several mm. There was no fracture for wet tests.

Table IV. Tests parameters and maximum number of cycles till fracture or end of test.

| Conditions | Air | Calf serum |
|---|---|---|
| Force/Frequency | 9 kN / 2 Hz | |
| Microseparation | 1.3 mm | |
| Number of cycles | 254,000 +/- 43,000 | No fracture after 800,000 cycles |

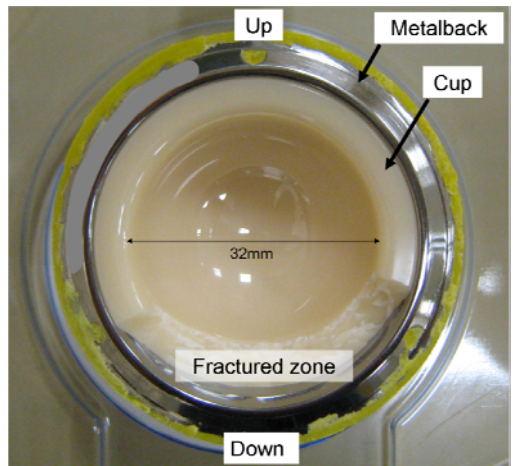

Figure 10. Alumina cup tested in dry conditions. Fracture at the lower rim.

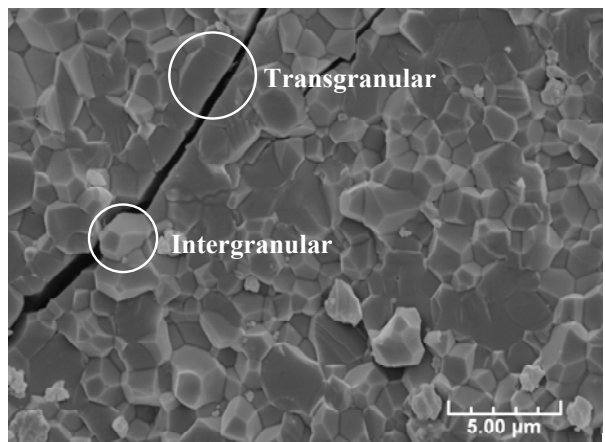

Figure 11. SEM image from alumina debris. Intergranular and transgranular fracture.

*Femoral Head*
   *Wear stripes*

Wear stripes were observed for all heads in the contact zone of the head with the rim of the cup, Figure 12a. There were no visible macroscopic cracks.

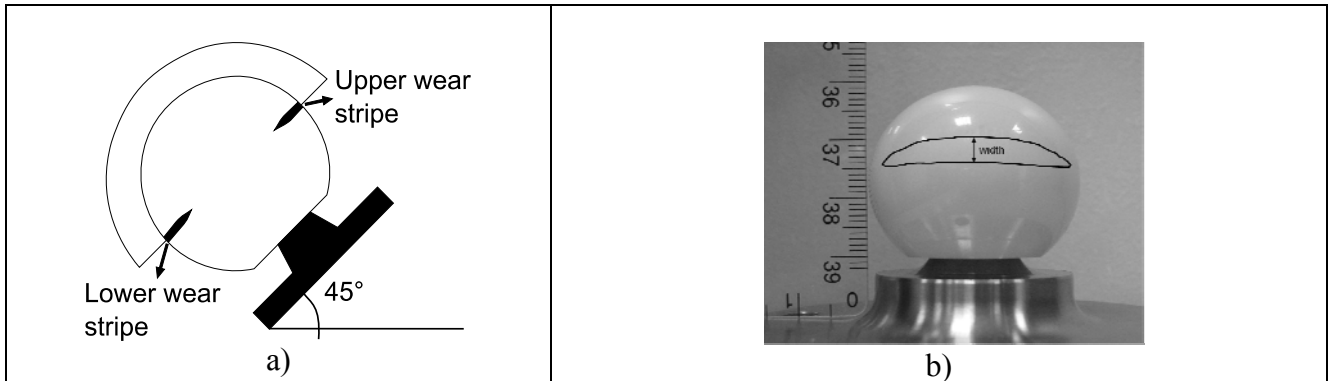

Figure 12. a) Location of wear stripes and b) Wear stripe on a head tested in dry conditions.

For dry tests, a wear stripe appears in first 50,000 cycles, while it appears only after 100,000 cycles in wet conditions. The location and general shape of the stripes were similar for shock machine and retrieved femoral heads[11]. Figure 12b shows wear stripes location. The width of the wear stripes was measured every 50,000 cycles for dry conditions and every 100,000 cycles for wet conditions, Figure 13. Since shape of the wear stripe exhibited on the head was not well defined at first cycles, first measurement of width was taken after 100,000 cycles for dry tests and after 250,000 cycles for wet tests.

Average width of wear stripe was strongly different for heads tested in dry and wet conditions; wear stripes were wider than 4mm in dry conditions while it was less than 2mm for wet conditions, Figure 13. However, at a macroscopic scale, the worn surfaces from the dry tests and wet tests were morphologically indistinguishable. Since wear stripes did not reached 4mm and had a small growth rate, tests were stopped at 800,000 cycles in wet conditions.

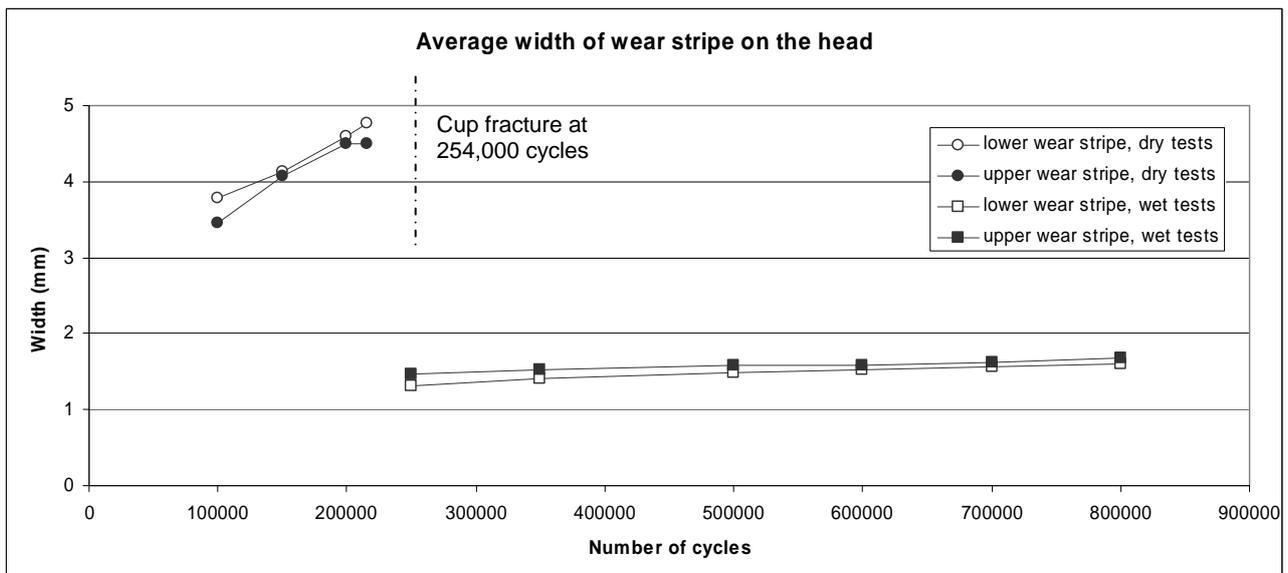

Figure 13. Width of wear stripes on the head for dry and wet tests.
Enlever lignes horizontales

In both dry and wet conditions there was an accelerate growth in the first 150,000 cycles; the slope is bigger for dry than the one for wet. After 200,000 cycles the width of the wear stripe of heads increased slowly. Figure 14 shows the average width of the wear stripes at the end of the tests.

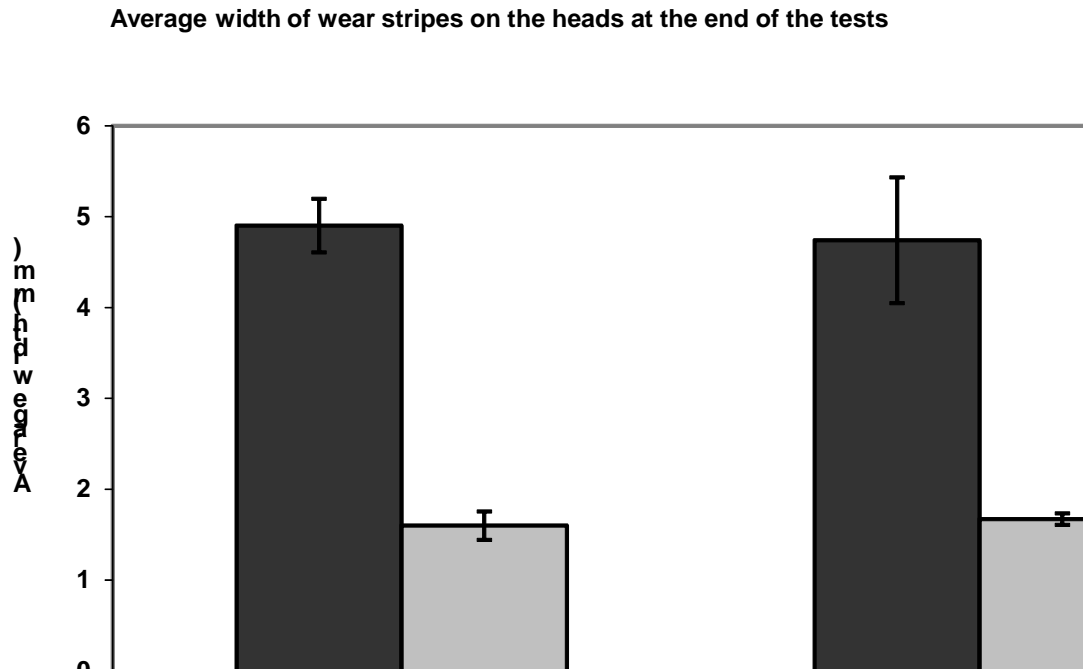

Figure 14. Width of wear stripes for heads in dry and wet conditions.
Problème avec la legende ?

*Roughness*

After the shock tests, heads were cut carefully and analysed by Atomic Force Microscopy (AFM, Software 6.13) to determine surface roughness evolution. A first order flatten filter was applied to suppress the effect of the spherical shape. 3D-roughness was of $9.1 \pm 5.1$ nm in unworn zones whereas it was measured to be $277.9 \pm 29.4$ nm in worn zones. Roughness in worn zones is 20 to 30-fold higher than in unworn zones. These values are comparable with retrieved heads[12]. A three dimensional reconstruction of the surface was made using Matlab®. Figure 15 shows surface for a worn and unworn zone from the same head.

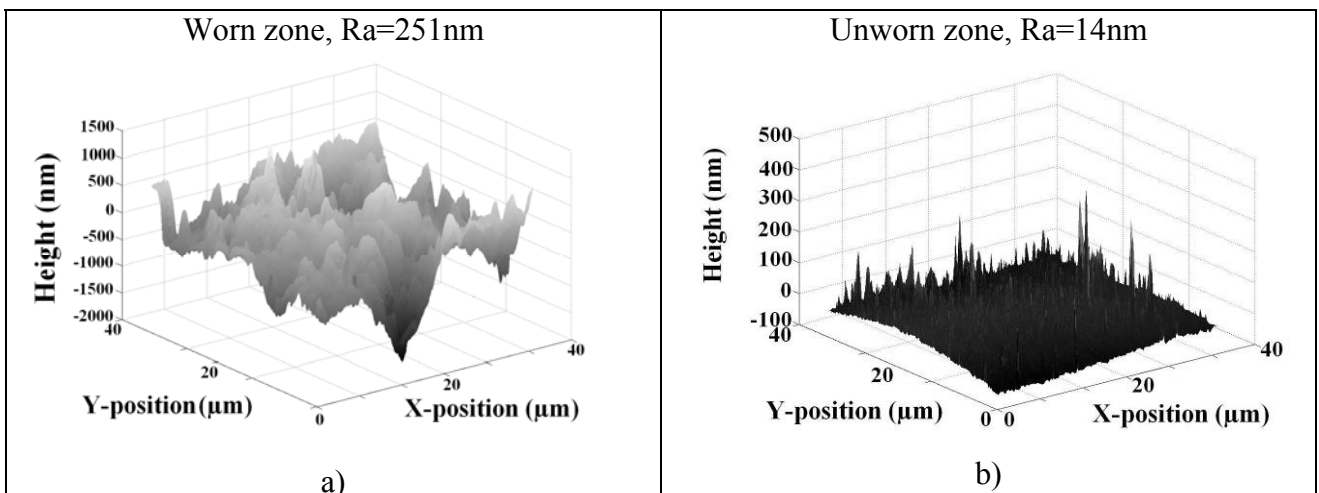

Figure 15. Three-dimensional reconstruction of a roughness profile for both a) worn and b) unworn zones studied by AFM.

Analyses by AFM showed a similar roughness for the upper and lower wear stripe, showed in Figure 16.

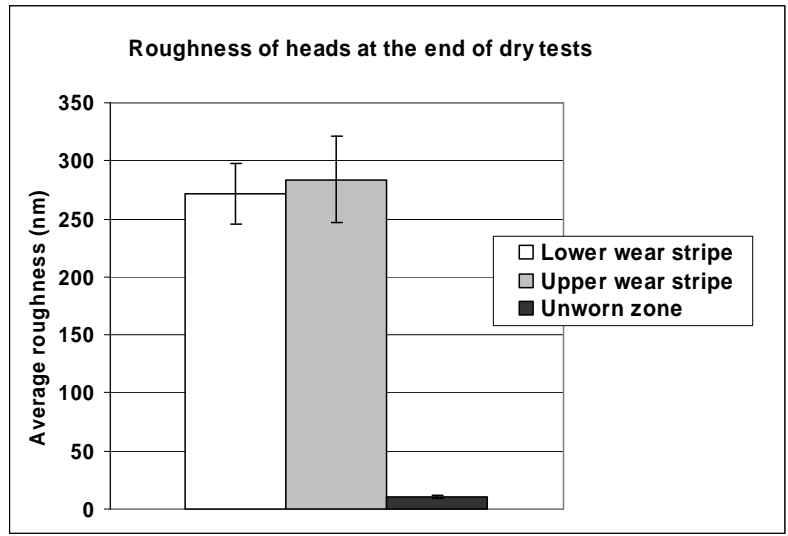

Figure 16. Wear stripe roughness from heads tested in dry conditions.

Enlever les lignes horizontales et agrandir le graphe.

## DISCUSSION

Movement of the cup after impact was studied using two LVDT sensors. The signal was modelled as an underdamped oscillator and data fit well. The k and c values depend on the mass of the actuator where the cup is fixed. The influence of those parameters in the stresses distribution on the cup should be studied in a finite element analysis.

An assembly of 32mm diameter cup and head manufactured of alumina was used in machine validation. Three types of signals were tested: sinus, triangle and gait signal. Several microseparation values were studied with increments of 0.5mm in the virtual distance g. From Figure 9, the working range of the shock machine was determined for gait cycle signal. An equation was then developed in order to calculate the relative distance, g, required to reach the force. Thus the shock device is calibrated for improving the microseparation and the specific degradation due the contact heel-ground. During the machine validation phase an equation of force as a function of the relative distance of the head was obtained. Relative distance, g, was then fixed to obtain a force of 9kN and 6 bearing surfaces were tested. Heads tested in both dry and wet conditions exhibited two wear stripes located in the zone of contact with the rim of the cup, at the upper and lower zones. Even tough the growth rate of wear stripes was higher at the beginning of the test, the width of wear stripes evolved linearly with the number of shocks. After shocking, the femoral head returns to its initial position in the vertical direction y, and cup oscillates until the next shock; the lower rim of the cup impacts the head, which could explain the fact that two wear stripes are observed on the head. Since the cup has a degree of freedom in the direction perpendicular to the force, the head might rebound into the cup leading to a sliding movement which might lead to pull out of the grains and an increasing width of the wear stripe. Moreover cups always fractured at the lower rim. It may be due to rebounds of the head inside the cup when head is going back to its initial position controlled by the force actuator.

Location of wear stripes corresponded well to those found *ex vivo* [6,12-18]. Nevertheless they are wider than those *in vivo*, which means a bigger worn surface and probably a wear rate more elevated. The number of shocks and the normal load are exacerbated in these investigated tests. Moreover the shocks frequency is two times higher than the one of the actual gait. It is the reason why the wear stripes are wider in these investigations. The highest force could be due to going down stairs. This study points out that this kind of shocks is relevant with the typical *in vivo* wear stripe. Due to the comparison between wet and dry conditions, the width of wear stripes, in the actual case, could be due to contact in both conditions. We investigated both conditions. One might suggest, for artificial hip joint, both conditions could occur for inducing the wear stripe in ceramic heads from implants. This could be explained by the fact that in physiological conditions, hyaluronic acid (HA) and albumin are responsible for the high viscosity and enhance boundary lubrication through adsorption on the joint material surfaces [19]. When using calf serum as lubricant, the lubrication regime could vary due to the degradation of proteins which could lead in a high wear rate because of an un-physiological protection effect against wear.

Several heads were sawed and five measurements of roughness were made by AFM. Difference between roughness of upper and lower wear stripes was not statistically significant. Compared to explants, wear stripes showed no significant differences; location, shape and roughness were in agreement with explants. Thus, despite the fact that lubrication conditions may vary from one patient to another one, the *ex vivo* degradation was observed in lab conditions when tests were carried out with calf serum.

Figure 11 shows that debris could be mostly about 1µm of diameter. Shocks induce these typical debris. This experimental result involves some expectations about the reactions of cells. Studies on the biological tolerance of alumina and zirconia nanosized powder on rats suggested that the inflammatory response is similar to those reported for polyethylene or metal particles [23]. Nevertheless, testing composites under shocks could lead to intergranular and transgranular fracture and smaller nanometric debris could be found, which might induce an inflammatory response.

Tests in hip simulator have shown that a small wear is reached for ceramic bearing surfaces [21]. Composites have a fracture toughness factor, $K_{IC}$, bigger than alumina and zirconia [22]. Further investigations in testing composites under shocks could give interesting results and it could be a validation phase for using such new materials in orthopaedic implants.

**CONCLUSIONS**

Based on the experimental results, the following conclusions are drawn related to the shock machine performance. A test device developed for testing hip prostheses has been tested. The shock machine allows studying materials behaviour before implantation and reproduce *in vitro* the *in vivo* wear rate. Indeed the device allows controlling microseparation between head and cup. The working range of the machine was determined for an alumina bearing surface of diameter 32mm. Thus the mechanical behaviour of the device is entirely determined thanks to experiments in various conditions. The mechanical impedance of the device is also determined and some modellings could be investigated in order to consider the influence of this one on shocks tests.

During shocks tests, as expected, a smaller degradation was observed in wet conditions when calf serum is used as lubricant. Width of wear stripes could be useful in estimating damage of the cup. Fracture occurred when the wear stripe on the head was wider than 4mm. Shock energy is very harmful in first contact head cup. Failure mechanism could be due to sliding after shocking. Moreover AFM analyses and wear stripes localisation are in agreement with *ex vivo* results. Since there was no significant difference between experimental results and retrieved heads, it can be inferred that this experimental device reproduces the actual degradation characteristics of the assembly head-cup made from ceramics. It is generally assumed that alumina has a high wear resistance. However, in view of the wear results from these tests, it can be noted that it is less wear resistant and brittle when subjected to shocks.

The point about this typical device is one might suggest that debris generation understanding could be improved with this device. Further analysis (morphological and statistical) should be carried out in order to confirm the presence of smaller wear debris. This experimental device will be used for testing new materials as composites alumina-zirconia. Finally enlarging the investigations field is an opportunity for other materials couples. Because the round ligament does not exist between femoral head and acetabulum, the shock mechanism is involved with metal-metal and metal-polymer (Ultra High Molecular Weight PolyEthylene). The tribowear problem of the artificial joint is constituted with shock degradation; one might suggest this mechanism should be examined with all materials couples.

Additional investigations will be carried out about the wear volume measurements thanks to optical profilometry. Moreover, one should pay attention on debris for characterizing them in terms of sizes and shapes. Thus the behaviour of bone cells, as osteoblasts or osteoclast, should be studied with debris coming from the shock experiments.

**ACKNOWLEDGEMENTS**

The authors want to acknowledge the Agence Nationale de la Recherche, ANR, for the financial support of the project 'Opt-Hip' ANR-07-MAPR-0014-02. The authors are grateful to A-M. Danna for AFM, L. Navarro for Matlab® investigations and N. Curt for the mechanical device.

**REFERENCES**

[1] Capello WN, D'Antonio JA, Feinberg JR, Manley MT, Naughton M. Ceramic-on-ceramic total hip arthroplasty: update. *The Journal of arthroplasty* 2008; **23** : 39-43, DOI: 10.1016/j.arth.2008.06.003.
[2] Murali R, Bonar SF, Kirsh G, Walter WK, Walter WL. Osteolysis in third-generation alumina ceramic-on-ceramic hip bearings with severe impingement and titanium metallosis. *Journal of arthroplasty* 2008; **23** : 1240.e13-9, DOI: 10.1016/j.arth.2007.10.020.
[3] Dennis DA, Komistek RD, Northcut EJ, Ochoa JA, Ritchie A. *In vivo* determination of hip joint separation and the forces generated due to impact loading conditions. *Journal of Biomechanics* 2001; **34** : 623-629, DOI: 10.1016/S0021-9290(00)00239-6.
[4] Hausselle J, Drapier S, Geringer J, Dursapt M, Stolarz J, Forest B. Modélisation de la croissance de défauts dans des cupules de prothèses de hanche en zircone soumises au phénomène de décoaptation. *Mécanique & Industries*, 2008 ; **9** : 153-158, DOI : 10.1051/meca:2008020.
[5] Bergmann G, Deuretzbacher G, Heller M, Graichen F, Rohlmann A, Strauss J, G.N. Duda. Hip contact forces and gait patterns from routine activities. *Biomechanics* 2001; **34** : 859–871, DOI: 10.1016/S0021-9290(01)00040-9.
[6] Nevelos J, Ingham E, Doyle C, Streicher R, Nevelos A, Walter W, Fisher J. Microseparation of the centers of alumina-alumina artificial hip joints during simulator testing produces clinically relevant wear rates and patterns. *The Journal of arthroplasty* 2000; **15** : 793-5. DOI: 10.1054/arth.2000.8100.
[7] Sariali E, Stewart T, Jin Z, Fisher J. Three-dimensional modeling of in vitro hip kinematics under micro-separation regime for ceramic on ceramic total hip prosthesis: an analysis of vibration and noise. *Journal of biomechanics* 2010; 43 : 326-333. DOI:10.1016/j.jbiomech.2009.08.031.
[8] Williams S, Jalali-Vahid D, Brockett C, Jin Z, Stone MH, Ingham E, Fisher J Effect of swing phase load on metal-on-metal hip lubrication, friction and wear. *Journal of biomechanics* 2006; 39 : 2274-2281. DOI:10.1016/j.jbiomech.2005.07.011.
[9] Stewart TD, Tipper JL, Insley G, Streicher RM, Ingham E, Fisher J. Severe wear and fracture of zirconia heads against alumina inserts in hip simulator studies with microseparation. The Journal of Arthroplasty 2003; 18 : 726-734. DOI:10.1016/S0883-5403(03)00204-3.
[10] Oates KMN, Krause WE, Jones RL, Colby RH . Rheopexy of synovial fluid and protein aggregation. *Journal of the Royal Society, Interface / the Royal Society*, 2006; **3** : 167-74. DOI: 10.1098/rsif.2005.0086.0086.
[11] Magnissalis EA, Xenakis TA, Zacharis C. Wear of retrieved ceramic THA components-Four matched pairs retrieved after 5-13 years in service. *J. Biomed. Mater. Res. Appl. Biomater* 2000; **158** : 593-598, DOI: 10.1002/jbm.1057.

[12] Nevelos JE, Ingham E, Doyle C, Fisher J, Nevelos AB. Analysis of retrieved alumina ceramics components from Mittelmeier total hip prostheses. *Biomaterials* 1999; **20** : 1833-1840, DOI: 10.1016/S0142-9612(99)00081-2

[13] Walter WL, Insley GM, Walter WK, Tuke MA. Edge loading in third generation alumina ceramic-on-ceramic bearings: stripe wear. *The Journal of Arthroplasty* 2004; **19** : 402-413, DOI: 10.1016/j.arth.2003.09.018.

[14] Manaka M, Clarke IC, Yamamoto K, Shishido T, Gustafson A, Imakiire A. Stripe wear rates in alumina THR-Comparison of microseparation simulator study with retrieved implants. *Journal of biomedical materials research. Part B, Applied biomaterials* 2004; **69** : 149-157, DOI: 10.1002/jbm.b.20033.

[15] Stewart TD, Tipper JL, Insley G, Streicher RM, Ingham E, Fisher J. Long-term wear of ceramic matrix composite materials for hip prostheses under severe swing phase microseparation. *Journal of biomedical materials research. Part B, Applied biomaterials* 2003; **66** : 567-573, DOI: 10.1002/jbm.b.10035.

[16] Shishido T, Yamamoto K, Tanaka S, Masaoka T, Clarke IC, Williams P. A Study for a retrieved implant of ceramic-on-ceramic total hip arthroplasty. *The Journal of arthroplasty* 2006; **21** : 294-298, DOI: 10.1016/j.arth.2005.05.025.

[17] Nevelos JE, Ingham E, Doyle C, Nevelos AB, Fisher J. Wear of HIPed and non-HIPed alumina-alumina hip joints under standard and severe simulator testing conditions. *Biomaterials*, 2001; **22** : 2191-7, DOI: 10.1016/S0142-9612(00)00361-6.

[18] Clarke IC, Green DD, Williams PA, Kubo K, Pezzotti G, Lombardi A, Turnbull A, Donaldson TK. Hip-simulator wear studies of an alumina-matrix composite (AMC) ceramic compared to retrieval studies of AMC balls with 1–7 years follow-up. *Wear*, 2009; **267** : 702-709, DOI: 10.1016/j.wear.2009.02.018.

[19] Tadmor R, Chen N, Israelachvili JN. Thin film rheology and lubricity of hyaluronic acid solutions at a normal physiological concentration. *Journal of biomedical materials research* 2002; **61** : 514-23, DOI: 10.1002/jbm.10215.

[20] Roualdes O, Duclos ME, Gutknecht D, Frappart L, Chevalier J, Hartmann DJ. In vitro and in vivo evaluation of an alumina-zirconia composite for arthroplasty applications. *Biomaterials* 2010; **31** : 2043-54, DOI: 10.1016/j.biomaterials.2009.11.107.

[21] Essner A, Sutton K, Wang A. Hip simulator wear comparison of metal-on-metal, ceramic-on-ceramic and crosslinked UHMWPE bearings. *Wear* 2005; **259** : 992-995, DOI: 10.1016/j.wear.2005.02.104.

[22] De Aza AH, Chevalier J, Fantozzi G, Schehl M, Torrecillas R. Crack growth resistance of alumina, zirconia and zirconia toughened alumina ceramics for joint prostheses. *Biomaterials* 2002; **23** : 937-945, DOI: 10.1016/S0142-9612(01)00206-X.

Toutes les corrections ne sont pas indiquées, je ne maîtrise pas encore la nouvelle version de Word 2011. Vous les vérifierez avec votre version d'origine.